\documentclass[twocolumn,pre,showpacs,amsmath,amssymb]{revtex4-1}
\usepackage{amsmath}
\usepackage{dcolumn}
\usepackage{wrapfig}
\usepackage{graphicx}
\usepackage{color}
\usepackage{graphics}
\usepackage{epsfig}
\usepackage{units}

\makeatletter
\begin{document}
\title{Inferring the origin of an epidemic with a dynamic message-passing algorithm}
\author{Andrey Y. Lokhov$^{1}$, Marc M\'ezard$^{1,2}$, Hiroki Ohta$^{1}$, and Lenka Zdeborov\'a$^{3}$} 
\affiliation{$^{1}$LPTMS, Universit\'e Paris-Sud and CNRS-UMR 8626, 91405 Orsay, France,}
\affiliation{$^{2}$Ecole Normale Sup\'erieure, 45 rue d'Ulm, 75005 Paris, France,}
\affiliation{$^{3}$IPhT, CEA Saclay and CNRS-URA 2306, 91191 Gif-sur-Yvette, France.}
\date{\today}

\begin{abstract}
We study the problem of estimating the origin of an epidemic outbreak:  given a contact network and a snapshot of epidemic spread at a certain time, determine the infection source.  This problem is important in different contexts of computer or social networks. Assuming that the epidemic spread follows the usual susceptible-infected-recovered  model, we introduce an inference algorithm based on dynamic message-passing equations and we show that it leads to significant improvement of performance compared to existing approaches. Importantly, this algorithm remains efficient in the case where the snapshot sees only a part of the network.
\end{abstract}


\maketitle

\section{Introduction}

Understanding and controlling the spread of epidemics on networks of contacts is an important task of today's science. It has far-reaching applications in mitigating the results of epidemics caused by infectious diseases, computer viruses, rumor spreading in social media and others. In the present article we address the problem of estimation of the origin of the epidemic outbreak (the so-called patient zero, or infection source - in what follows, these terms 
are used alternately): given a contact network and a snapshot of epidemic spread at a certain time, determine the infection source. Information about the origin could be extremely useful to reduce or prevent future outbreaks. Whereas the dynamics and the prediction of epidemic spreading in networks have attracted a considerable number of works, for a review see~\cite{Murray89,Hethcote00,BoccalettiaLatora06}, the  problem of estimating the epidemic origin has been mathematically formulated only recently~\cite{ShahZaman10}, followed by a burst of research on this practically important problem~\cite{CominCosta11,ShahZaman11,ZhuYing12,PrakashVreeken12,FioritiChinnici12,PintoThiran12,DongZhang13}. In order to make the estimation of the origin of spreading a well-defined problem we need to have some knowledge about the spreading mechanism. We shall adopt here the same framework as in existing works, namely we assume that the epidemic spread follows the widely used susceptible-infected-recovered (SIR) model or some of its special cases~\cite{KermackMcKendrick27,anderson1991infectious}. 

The stochastic nature of infection propagation makes the estimation of the epidemic origin intrinsically hard: indeed, different initial conditions can lead to the same configuration at the observation time. Finding an estimator that locates the most probable origin, given observed configuration, is in general computationally intractable, except in very special cases such as the case where the contact network is a line or a regular tree~\cite{ShahZaman10,ShahZaman11,DongZhang13}. The methods that have been studied in the existing works are mostly based on various kinds of graph-centrality measures. Examples include the distance centrality or the Jordan center of a graph~\cite{ShahZaman10,CominCosta11,ShahZaman11,ZhuYing12}. The problem was generalized to estimating a set of epidemic origins using spectral methods in~\cite{PrakashVreeken12,FioritiChinnici12}. Another line of approach uses more detailed information about the epidemic than just a snapshot at a given time~\cite{PintoThiran12}. 
Note, however, that all the present methods are limited, for instance none of them makes an efficient use of the information about the nodes to which the epidemic did not spread. 

In this paper we introduce a new algorithm for the estimation of the origin of an SIR epidemic from the knowledge of the network and the snapshot of some nodes at a certain time. 
Our algorithm estimates the probability that the observed snapshot resulted from a given patient zero in a way which is crucially different from existing approaches. 
For every possible origin of the epidemic, we use a fast dynamic message-passing method to estimate the probability that a given node in the network was in the observed state ($S$, $I$ or $R$). We then use a mean-field-like approximation to compute the probability of the observed snapshot as a product of the marginal probabilities. We finally rank the possible origins according to that probability. 

The dynamic message-passing (DMP) algorithm that we use in order to estimate the probability of a given node to be in a given state is interesting in itself. It belongs to the class of message-passing algorithms that includes the standard belief propagation (BP) method, also known in different fields as cavity method or sum-product equations~\cite{YedidiaFreeman03,MezardMontanari07}. BP is a distributed technique that allows to estimate marginal probability distributions in problems on factor graphs and networks, and appeared to be very successful when applied to Bayesian networks \cite{Pearl88}, error-correcting codes \cite{Gallager62} and optimization problems \cite{MezardParisi02}. The BP equations are derived from the Boltzmann-Gibbs distribution under the assumption that the marginals defined on an auxiliary cavity graph (a graph with a removed node) are uncorrelated, which is exact if the underlying network is a tree (for a general discussion, see \cite{MezardMontanari07}). From a numerical point of view, the solution of the BP equations is obtained by iteration until convergence.

The DMP equations can be derived be generalizing BP to dynamic problems, using as variables in the corresponding graphical model the whole time trajectories of a given node, see e.g. \cite{KanoriaMontanari11,neri2009cavity}. For general dynamics, the complexity of the equations increases exponentially with time, making it impossible to solve the dynamic BP equation for the whole trajectory except for only few time steps. However, crucial simplifications occur for the models with irreversible dynamics, such as the random field Ising model~\cite{OhtaSasa10} and the SIR model, considered in this paper. Indeed, the time trajectories in these models can be fully parametrized with only few flipping times, leading to important simplification of the corresponding BP equations on trajectories. As a result, they can be rewritten in terms of closed DMP equations, using dynamic variables that appear to be the weighted sums of messages of the dynamic BP equations. In this work, we present a more straightforward derivation that makes use of arguments similar to those used in the cavity method \cite{MezardMontanari07}.      

A precursor of DMP equations appeared in~\cite{KarrerNewman10b} in a form averaged over initial conditions, which does not lend itself to algorithmic use. Here we derive and use the  DMP on a given network for given initial conditions. If averaged also over the graph ensemble, it can be used to obtain the asymptotically exact dynamic equations of~\cite{Volz08,Miller11} for SIR, or those of~\cite{OhtaSasa10} for avalanches in the random field Ising model. Note that although DMP bares some similarity with BP, it is crucially different in several aspects: it is not directly derived from a Boltzmann-like probability distribution, and it does not need to be iterated until convergence; instead the iteration time corresponds directly to the real time in the associated SIR dynamics. A nice property that DMP shares with BP is that it gives exact results if the contact network is a tree. We use it here as an approximation for loopy-but-sparse contact networks in the same way that BP is commonly used with success in equilibrium studies of such networks.

We test our algorithm on synthetic spreading data and show that it performs better than existing approaches (except for a special region of parameters where the Jordan center is on average better). The algorithm is very robust, for instance it remains efficient even in the case where the states of only a fraction of nodes in the network are observed. From our tests we also identify a range of parameters for which the estimation of the origin of epidemic spreading is relatively easy, and a region where this problem is hard. Hence, our dataset can also serve as a test-bed for new approaches.

\section{SIR model and dynamic message-passing equations}

\subsection{Spreading model}

The mathematical modeling of epidemic spreading is a subject of growing interest because of its importance for practical applications, such as analysis, evaluation and prevention of consequences of epidemiological processes. Percolation-like processes have been addressed in a number of physics studies, in particular aiming at understanding the role of the network topology on the spreading results. The most popular and studied epidemiological models are susceptible-infected-susceptible (SIS) and susceptible-infected-recovered (SIR) models. The SIS model is used to model endemic diseases that can be maintained in a population for a long time because of the reinfection of individuals. In the SIR model, the infection can not persist indefinitely due to depletion of susceptible agents, and the quantity of interest is typically the fraction of population touched by the infection. The general properties and the phase diagram of these models on random networks were studied in many works, see e.g.~\cite{BoccalettiaLatora06} and references therein. In this paper, we focus on the SIR model, corresponding to the irreversible dynamic process where the nodes that catch the infection ultimately get either immunized or dead. The real cases that fall into this category include diseases that confer immunity to their survivors, or computer viruses in a setting of a permanent virus-checking against attacks of the same virus in a computer network.

The typical assumptions in the studies of the SIR dynamics on networks include the uniformity of the infection and recovery probabilities and the mass-action mixing hypothesis, i.e. an assumption that in principle each pair of individuals can interact, ignoring the actual topology of the physical contacts. These assumptions allow to write simple ``naive'' mean field differential equations on the densities of susceptible, infected and recovered nodes in the population, providing for a qualitative understanding of mechanisms and thresholds of epidemic spreading and a rough fitting of some real epidemic data \cite{KermackMcKendrick27,anderson1991infectious}. However, these assumptions are obviously unrealistic since they do not account for heterogeneities in contacts and transmission probabilities. A number of recent investigations addressed these issues by considering more accurate settings, e.g. using random networks, however, averaging over the graph ensembles or initial conditions (for reviews see \cite{BoccalettiaLatora06,Newman03,KarrerNewman10b}). Still, most of the studies on random graphs are limited to numerical simulations or to the use of naive mean field equations, which might be a crude approximation for some applications. In this paper, we study the SIR model on a given graph. The dynamic equations that we use are exact for locally tree-like networks; for real-world problems they often provide a good approximation, allowing a better determination of the infection source. Throughout this work we study a static network of interacting individuals, although dynamically changing networks can also be considered within our approach, see discussion below.

The SIR model is defined as follows. Let $G\equiv(V,E)$ be a connected undirected graph containing $N$ nodes defined by the set of vertices $V$ and the set of edges $E$. Each node $i \in V$ at discrete time $t$ can be in one of three states $q_{i}(t)$: susceptible, $q_{i}(t)=S$, infected, $q_{i}(t)=I$, or recovered, $q_{i}(t)=R$. At each time step, an infected node $i$ will recover with probability $\mu_i$, and a susceptible node $i$ will become infected with probability $1-\prod_{k\in\partial i}(1-\lambda_{ki}\delta_{q_{k}(t),I})$, where $\partial i$ is the set of neighbors of node $i$, and $\lambda_{ki}$ measures the efficiency of spread from node $k$ to node $i$. The recovered nodes never change their state. We assume that the graph $G$ and parameters $\lambda_{ij}$, $\mu_{i}$ are known (or have been already inferred).

\subsection{Dynamic message-passing equations}

Let us derive the dynamic message-passing equations (DMP) for the SIR model that are used later in the inference algorithm. In particular, we will show that the probabilities of being susceptible/infected/recovered at a given time $t$ as provided by the DMP equations are exact for all initial conditions and every realization of the transmission and recovery probabilities $\lambda_{ij}$ and $\mu_i$ if the graph of contacts is a tree. We define $P_{S}^{i}(t)$, $P_{I}^{i}(t)$ and $P_{R}^{i}(t)$ as the marginal probabilities that $q_{i}(t) = S$, $q_{i}(t) = I$ and $q_{i}(t) = R$. These marginals sum to one and thus
\begin{align}
P_I^{i}(t+1)=1-P_S^{i}(t+1)-P_R^{i}(t+1).
\label{eq:I_dynamics}
\end{align}
Since the recovery process from state $I$ to state $R$ is independent of neighbors, we have
\begin{align}
P_R^{i}(t+1)=P_R^{i}(t)+\mu_{i}P_{I}^{i}(t).
\label{eq:R_dynamics}
\end{align}

The epidemic process on a graph can be interpreted as the propagation of infection signals from infected to susceptible nodes. The infection signal $d^{i \rightarrow j}(t)$ is defined as a random variable which is equal to one with probability $\delta_{q_{i}(t-1),I}\lambda_{ij}$, and equal to zero otherwise. Consider an auxiliary dynamics $D_{j}$ where node $j$ receives infection signals, but ignores them and thus is fixed to the $S$ state at all times. Since the infection cannot propagate through node $j$ in this dynamic setting, different graph branches rooted at node $j$ become independent if the underlying graph is a tree. This is the natural generalization of the cavity method used for deriving BP (see~\cite{MezardMontanari07}) to dynamic processes. Notice that the auxiliary dynamics $D_{j}$ is identical to the original dynamics $D$ for all times such that $q_{j}(t)=S$. We also define an auxiliary dynamics $D_{ij}$ in which the state of a pair of neighboring nodes $i$ and $j$ is always $S$.

In order to obtain a closed system of message-passing equations, we write the remaining update rules in terms of three kinds of cavity messages, defined as follows. We first define the message $\theta^{k \rightarrow i}(t)$ as the probability that the infection signal has not been passed from node $k$ to node $i$ up to time $t$ in the dynamics $D_{i}$:
\begin{align}
\theta^{k \rightarrow i}(t) = \text{Prob}^{D_{i}} \left( \sum_{t'=0}^t d^{k \rightarrow i}(t')=0 \right).
\label{eq:theta_definition}
\end{align}
The quantity $\phi^{k \rightarrow i}(t)$ is the probability that the infection signal has not been passed from node $k$ to node $i$ up to time $t$ in the dynamics $D_{i}$ and that node $k$ is in the state $I$ at time $t$:
\begin{align}
\phi^{k \rightarrow i}(t) = \text{Prob}^{D_{i}} \left( \sum_{t'=0}^{t} d^{k \rightarrow i}(t')=0, \, q_{k}(t)=I \right).
\end{align}
Finally, $P_{S}^{k \rightarrow i}(t)$ is the probability that node $k$ is in the state $S$ at time $t$ in the dynamics $D_{i}$:
\begin{align}
P_{S}^{k \rightarrow i}(t) = \text{Prob}^{D_{i}} \left( q_{k}(t)=S \right).
\end{align}
In what follows, we prove that
\begin{align}
P_S^{i \rightarrow j}(t+1)=P_S^{i}(0)\prod_{k\in\partial i \backslash j}\theta^{k \rightarrow i}(t+1),
\label{eq:S_message_dynamics}
\end{align}
where $\partial i \backslash j$ means the set of neighbors of $i$ excluding $j$. Indeed, by definition
\begin{align}
\notag
P_S^{i \rightarrow j}(t+ & 1) = \text{Prob}^{D_{j}} \left( q_{i}(t+1)=S \right)
\\
& = P_S^{i}(0) \, \text{Prob}^{D_{j}} \left( \sum_{k\in\partial i \backslash j} \sum_{t'=0}^{t+1} d^{k \rightarrow i}(t')  \right).
\end{align}
Since the auxiliary dynamics $D_{ij}$ coincides with dynamics $D_{j}$ as long as node $i$ is in the $S$ state, we can write
\begin{align}
P_S^{i \rightarrow j}(t+1) =
P_S^{i}(0) \, \text{Prob}^{D_{ij}} \left( \sum_{k\in\partial i \backslash j} \sum_{t'=0}^{t+1} d^{k \rightarrow i}(t')  \right).
\end{align}
Since different branches of the graph containing nodes $k\in\partial i \backslash j$ are connected only through node $i$, they are independent of each other, hence
\begin{align}
P_S^{i \rightarrow j}(t+1) =
P_S^{i}(0) \prod_{k\in\partial i \backslash j} \text{Prob}^{D_{ij}} \left( \sum_{t'=0}^{t+1} d^{k \rightarrow i}(t')  \right).
\end{align}
Moreover, for nodes $k\in\partial i \backslash j$, the dynamics $D_{ij}$ is equivalent to the dynamics $D_{i}$, so we can replace $D_{ij}$ by $D_{i}$ in the last expression and hence, using the definition (\ref{eq:theta_definition}), we obtain equation (\ref{eq:S_message_dynamics}).
We complete the updating rules by writing the equations for $\theta^{k \rightarrow i}(t)$ and $\phi^{k \rightarrow i}(t)$. The only way in which $\theta^{k \rightarrow i}(t)$ can decrease is by actually transmitting the infection signal from node $k$ to node $i$, and this happens with probability $\lambda_{ki}$ multiplied by the probability that node $k$ was infected, so we have
\begin{align}
\theta^{k \rightarrow i}(t+1)-\theta^{k \rightarrow i}(t) = -\lambda_{ki}\phi^{k \rightarrow i}(t).
\label{eq:theta_dynamics}
\end{align}
The change for $\phi^{k \rightarrow i}(t)$ at each time step comes from three different possibilities: 
either node $k$ actually sends the infection signal to node $i$ 
(with probability $\lambda_{ki}$), 
either it recovers (with probability $\mu_{k}$), or it switches to $I$ at this time step, being previously 
in the $S$ state (this happens with probability $S^{i \rightarrow j}(t-1)-S^{i \rightarrow j}(t)$):

\begin{align}
\notag
\phi^{k \rightarrow i}(t)- \phi^{k \rightarrow i}(t-&1) =
-\lambda_{ki}\phi^{k \rightarrow i}(t-1)
\\
\notag
-\mu_{k}\phi^{k \rightarrow i}(t&-1)
+\lambda_{ki}\mu_{k}\phi^{k \rightarrow i}(t-1)
\\
&+S^{k \rightarrow i}(t-1)-S^{k \rightarrow i}(t).
\label{eq:phi_dynamics}
\end{align}
The third compensation term on the right-hand side of the previous equation has been introduced in order to avoid double-counting in the situation when 
node $k$ transmits the infection and recovers at the same time step.

This completes the update rules for cavity messages. These equations can be iterated in time starting from initial conditions for cavity messages:
\begin{eqnarray}
&\theta^{i \rightarrow j}(0)=1,\label{eq:initial_conditions1}
\\
&\phi^{i \rightarrow j}(0)=\delta_{q_i(0),I}.\label{eq:initial_conditions2}
\label{eq:initial_conditions} 
\end{eqnarray}
The marginal probability in the original dynamics $D$ is obtained by including all the neighbor nodes $k\in\partial i$ in eq.~(\ref{eq:S_message_dynamics}):
\begin{align}
P_S^{i}(t+1)=P_S^{i}(0)\prod_{k\in\partial i}\theta^{k \rightarrow i}(t+1).
\label{eq:S_dynamics} 
\end{align}
Let us summarize the closed set of recursion rules, given by the combination of (\ref{eq:I_dynamics}, \ref{eq:R_dynamics}, \ref{eq:S_message_dynamics}, \ref{eq:theta_dynamics}, \ref{eq:phi_dynamics}, \ref{eq:S_dynamics}):
\begin{align}
 P_S^{i \rightarrow j}(t+1)=P_S^{i}(0)\prod_{k\in\partial i \backslash j}&\theta^{k \rightarrow i}(t+1), \label{eq:SIRequations:Sc}
\\
\theta^{k \rightarrow i}(t+1)-\theta^{k \rightarrow i}(t) =
-&\lambda_{ki}\phi^{k \rightarrow i}(t),
\\
\notag 
\phi^{k \rightarrow i}(t)=(1-\lambda_{ki})(1-\mu_{k}&)\phi^{k \rightarrow i}(t-1)
\\
 -[P_S&^{k \rightarrow i}(t)-P_S^{k \rightarrow i}(t-1)].
\end{align}
The marginal probabilities that node $i$ is in a given state at time $t$ are then given as
\begin{align}
& P_S^{i }(t+1)=P_S^{i}(0)\prod_{k\in\partial i}\theta^{k \rightarrow i}(t+1)\, ,\label{eq:SIRequations:S}
\\
& P_R^{i}(t+1)=P_R^{i}(t)+\mu_{i}P_{I}^{i}(t)\, ,\label{eq:SIRequations:R}
\\
& P_I^{i}(t+1)=1-P_S^{i}(t+1)-P_R^{i}(t+1)\, .\label{eq:SIRequations:I}
\end{align}
Together with the initial conditions (\ref{eq:initial_conditions1}-\ref{eq:initial_conditions2}), these equations give the exact values of marginal probabilities $P_{S}^{i}(t)$, $P_{I}^{i}(t)$ and $P_{R}^{i}(t)$ on a tree graph. The algorithmic complexity of DMP equations for a given vertex $i$ is $O(tNc)$, where $c$ is the average degree of the graph.

It should be noted that equations reminiscent of (\ref{eq:SIRequations:Sc}-\ref{eq:SIRequations:I}) were first derived in \cite{KarrerNewman10b}. The authors of \cite{KarrerNewman10b} treated a more general SIR model where the transmission and recovery distributions are non-exponential. For this more general case, no easily tractable form of the DMP is known (by this we mean a Markovian form of the DMP, where  the probabilities at time $t$ give the probabilities at time $t+1$ via a set of simple closed equations). The equations in~\cite{KarrerNewman10b} were instead written in a convolutional form that is rather complicated for numerical resolution. 
The authors noticed that when recovery and transmission rates are constant, the equations simplify, but did not write a version of the equations that is applicable on a given graph for a given initial condition (actually they only wrote equations averaged over a set of initial conditions). Hence we find it useful to provide the derivation of the DMP on a single graph in their simple iterative form.

For the purpose of this paper we use the DMP on a single instance of the contact network for a given initial condition. However, if an ensemble of initial conditions is given as well as an ensemble of random graphs with a given probability distribution then one can write differential equations for the fraction of nodes that are susceptible/infected/recovered at a given time. These equations were first derived by \cite{Volz08} and appeared also in \cite{KarrerNewman10b} and \cite{Miller11}. One should not confuse these averaged DMP equations with the ``naive''  mean field equations that are often written for the SIR model under the assumption of perfect mixing, as discussed previously. Whereas the naive mean field equations provide only a very crude approximation for the real probabilities, the equations of \cite{Volz08,Miller11} are exact in the thermodynamic limit, $N\to \infty$, as long as, in the random graph ensemble, the probability that a randomly-chosen node belongs to a finite-length loop goes to zero in the large graph-size limit.

\section{Inference of epidemic origin and DMP algorithm}

\begin{figure}[!th]
\vspace{-2mm}
\includegraphics[width=0.5\columnwidth]{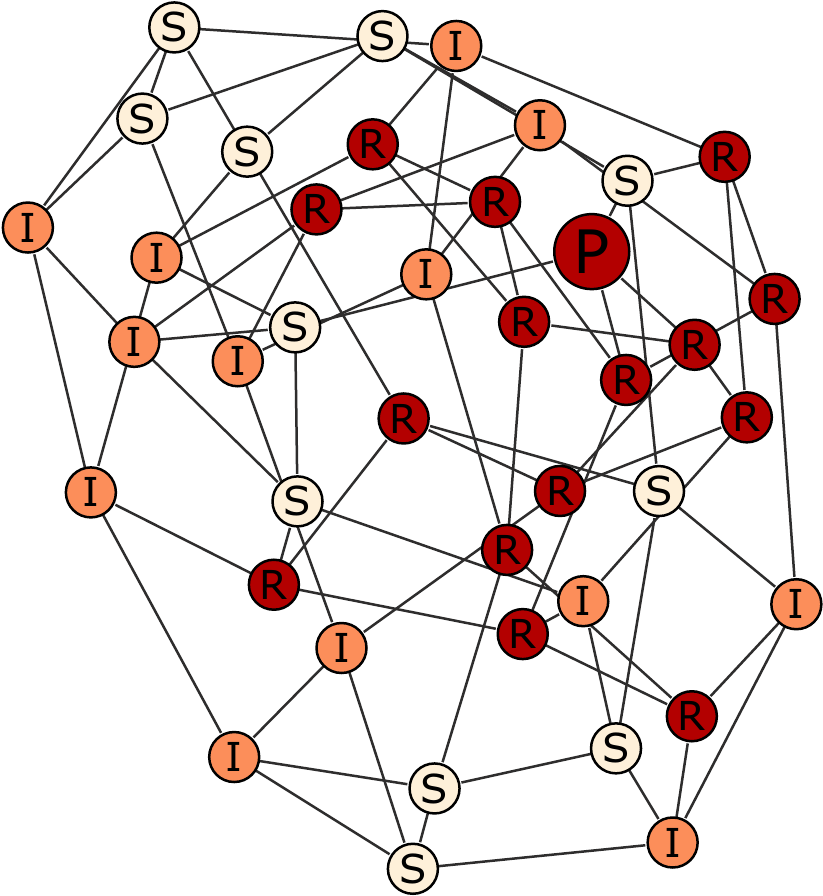}
\caption{(color online) An example of a single instance of the inference problem on a random regular graph of degree $c=4$ with $N=40$ nodes. The patient zero is labeled by $P$ and appears in the state $R$ in the snapshot. The epidemic is generated for $\lambda=0.5$ and $\mu=0.5$, the snapshot is represented at time $t_{0}=5$.
}
\label{fig:singleinstance}
\end{figure}

To define the problem of estimation of the epidemic origin, we consider the case where, at initial time $t=0$, only one node is infected (the ``patient zero'', $i_0$), and all others nodes are susceptible. After $t_0>0$ time steps  ($t_{0}$  is in general unknown), we observe the state of a set of nodes $\mathcal{O}\subset V$, and the task is to estimate the location of patient zero based on this snapshot, see Fig.~\ref{fig:singleinstance}.

\begin{figure}[!th]
\includegraphics[width=0.97\columnwidth]{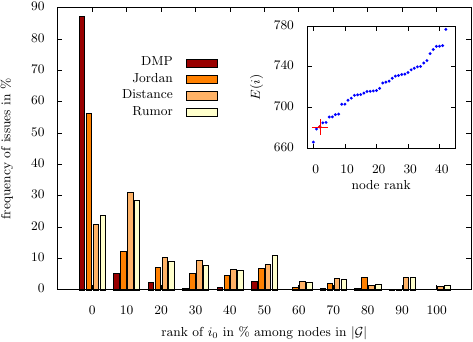}
\caption{(color online) A test of inference of the epidemic origin on random regular graphs of degree $c=4$, size $N=1000$. Inset: An epidemic is generated with recovery probability $\mu=1$, transmission probability $\lambda=0.6$, a snapshot of all the nodes is taken at time $t_0=8$
 (in this figure we assume we know the value of $t_0$), $242$  nodes are observed to be in the $I$ or $R$ state. The dynamic message-passing is used to compute the energy of every node. This energy is finite for $43$ nodes; it is plotted as a function of their rank $r$. The true patient zero is marked by a red cross, and its rank is $r(i_0)=2$ in this case. Main figure: an epidemic generated with $\mu=1$, $\lambda=0.5$, $t_0=5$. The histogram (over $1000$ random instances) of the normalized rank (i.e. the rank divided by the number of $R$ or $I$ nodes in the snapshot) of the true patient zero is plotted for the dynamic message-passing (DMP) inference, as well as for the distance, rumor and Jordan centrality measures. 
}
\label{fig:rankonxi}
\vspace{-2mm}
\end{figure} 

Let us briefly explain two existing algorithms~\cite{ShahZaman10,ShahZaman11,ZhuYing12} that we will use as benchmarks. The authors of~\cite{ShahZaman10,ShahZaman11,ZhuYing12} considered only the case when all the nodes were observed, $\mathcal{O}=V$. In appendix~\ref{app:centralities} we propose a generalisation of these algorithms to a more general case. 
The most basic measure for node $i$ to be the epidemic origin is the distance centrality $D(i)$ which we define as
$D(i)\equiv\sum_{j\in\mathcal{G}}d(i,j)\left(\delta_{q_j,I}+\delta_{q_j,R}/\mu_j\right)$, where the graph $\mathcal{G}$ is a connected component of the original graph $G$ containing all infected and recovered nodes and only them, and $d(i,j)$ is the shortest path between node $i$ and node $j$ on the graph $\mathcal{G}$. The ad-hoc factor $1/\mu_j$ is introduced to distinguish recovered nodes that for small $\mu_j$ tend to be closer to the epidemic origin. In the existing works this factor was not present, because~\cite{ShahZaman10,ShahZaman11} treated only the SI model, and~\cite{ZhuYing12} considered that susceptible and recovered nodes are indistinguishable. The authors of~\cite{ShahZaman10,ShahZaman11} suggested a ``rumor centrality'' estimator and showed that, for tree graphs, the rumor centrality and the distance centrality coincide.
Another simple but well-performing estimator, Jordan centrality $J(i)$, was proposed in~\cite{ZhuYing12} and corresponds to a node minimizing the maximum distance to other infected and recovered nodes:
$J(i)\equiv\max_{j\in\mathcal{G}}d(i,j)$.
A node $i$ where  $J(i)$ is minimal is known as a `Jordan center' of $\mathcal{G}$ in the graph theory literature. Note that in~\cite{ZhuYing12} the Jordan center of only the infected notes was used, hence our implementation uses more information. 

\begin{figure*}[!ht]
\begin{center}
\includegraphics[width=2\columnwidth]{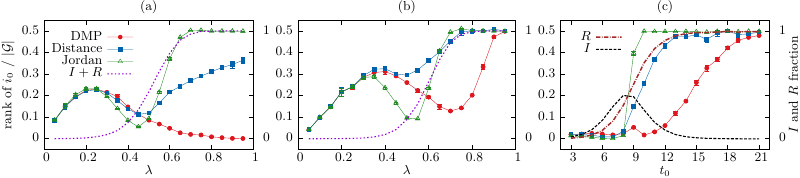}
\caption{(color online) Average rank of the true epidemic origin on random regular graphs of size  $N=1000$ with degree $c=4$. Each data point is averaged over $1000$ instances. The plots (a) and (b) represent the dependences of the average rank on the infection rate $\lambda$, for the snapshot time 
$t_0=10$ and recovery probability $\mu$: (a) $\mu=0.5$ and (b) $\mu=1$. In this figure $t_0$ is inferred by the algorithm.
 The DMP estimator (red circles) is compared to the Jordan centrality (green triangles) and the distance centrality (blue squares) estimators. The dotted line shows the average fraction of nodes that were infected or recovered in the snapshot, $|{\cal G}|/N$, we use this number to normalize the ranks of the epidemic origin. Plot (c) shows the dependence on the snapshot time $t_{0}$ for $\lambda=0.7$ and $\mu=0.5$. The dashed line is the average fraction of nodes that were infected and the dash-dotted line   is the average fraction of nodes that were recovered in the snapshot; both are normalized to $N$.}
\label{fig:2}
\end{center}
\vspace{-5mm}
\end{figure*}

\begin{figure*}[!ht]
\begin{center}
\includegraphics[width=1.86\columnwidth]{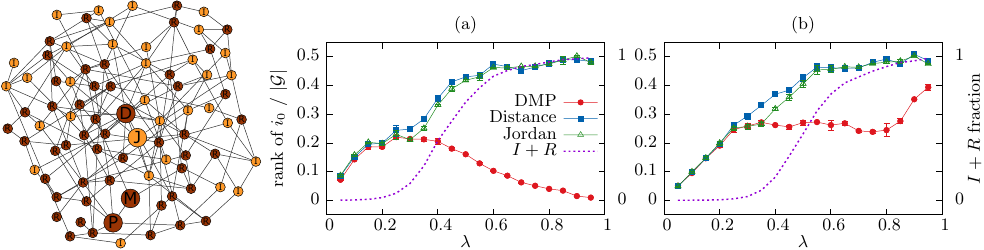}
\caption{(color online) Left: An instance of inference problem on the Erd\"os-R\'enyi graph with average degree $\langle c \rangle=4$ and $N=84$. The epidemic is generated for $\lambda=0.7$ and $\mu=0.5$. In this example, only infected (light) and recovered (dark) nodes are present in the snapshot at time $t_{0}=5$. The true patient zero is labeled by $P$, the best-ranked nodes for DMP ($M$), Jordan ($J$) and distance ($D$) centralities are at distances 1, 2 and 3 from $P$, correspondingly. Right: Average rank of the true epidemic origin on Erd\"os-R\'enyi graphs of size  $N\simeq1000$ with average degree $\langle c \rangle=4$. Each data point is averaged over $1000$ instances. 
The snapshot time $t_{0}$ (assumed to be known) and recovery probability $\mu$ are: (a) $t_{0}=10$, $\mu=0.5$ and (b) $t_{0}=10$, $\mu=1$. The dotted line shows the average fraction of nodes that were infected or recovered in the snapshot, $|{\cal G}|/N$, we use this number to normalize the ranks of the epidemic origin.}
\label{fig:ER}
\end{center}
\vspace{-5mm}
\end{figure*}   

The core of the algorithm proposed in the present work is DMP, explained in the previous section, which provides an estimate of the probabilities  $P_S^j(t,i_0)$ (respectively $P_I^j(t,i_0)$, $P_R^j(t,i_0)$) that a node $j$ is in each of the three states $S$, $I$, $R$, at time $t$, for a given patient zero $i_0$.
Let us first assume that the time $t_0$ is known. With the use of Bayes rule, the probability that node $i$ is the patient zero given the observed states is proportional to the joint probability of observed states given the patient zero, 
$P(i|\mathcal{O})\sim P(\mathcal{O}|i)$. We also define an energy-like function of every node $E(i)\equiv-\log{P(\mathcal{O}|i)}$, such that nodes with lower energy are more likely to be the infection source. If one were able to compute $P(\mathcal{O}|i)$ exactly, finding $i$ which minimizes $E(i)$ would be an optimal inference scheme of the patient zero. As there is no tractable way to compute exactly the joint probability of the observations,  we approximate it using a mean-field-type approach as a product of the marginal probabilities provided by the dynamic message-passing
\begin{equation}
P({\cal O} | i) \simeq  \hspace{-3mm}  \prod_{{k \in {\cal O} \atop q_{k}(t_0)=S}}   \hspace{-3mm}   P_S^{k}(t,i) \hspace{-3mm}  \prod_{{l \in {\cal O}  \atop q_{l}(t_0)=I}}  \hspace{-2mm}     P_I^{l}(t,i) \hspace{-3mm}  \prod_{{m \in {\cal O}  \atop q_{m}(t_0)=R}}   \hspace{-3mm}  P_R^{m}(t,i) \, .
\label{eq:CostFunction}
\end{equation}
To estimate the value of $t_0$, we compute the energy $E(i,t)$ for different possible values $t$, and choose the value that maximizes the ``partition function'' $Z(t)\equiv\sum_{i}e^{-E(i,t)}$. As mentioned previously, the algorithmic complexity for computing the energy $E(i)$ of a given vertex $i$ (and therefore the probability that it is the epidemic origin) is $O(t_{0}Nc)$, where $c$ is the average degree of the graph.

\section{Performance of inference algorithms}

We first test our algorithm on random regular graphs, i.e. random graphs drawn uniformly from the set of graphs where every node has degree $c$. 
In all the simulations we consider uniform transmission and recovery probabilities 
$\lambda_{ij}=\lambda$ and $\mu_i=\mu$.

In the first example, inset of Fig.~\ref{fig:rankonxi}, we plot the energy $E(i)$ resulting from the dynamic message-passing of the nodes for which the probability of being the epidemic origin is finite according to our algorithm, we order the nodes according to the energy value. The true epidemic origin is marked with a red cross. We define the rank of candidates for the epidemic origin to be its position in this ranking (the lowest energy node having rank $0$). The main graph of Fig.~\ref{fig:rankonxi} shows the histogram of normalized ranks (i.e. the rank divided by the total number of nodes that were observed as recovered or infected) of the true epidemic origin as obtained from our DMP inference algorithm, compared to the rankings obtained by distance, rumor and Jordan centralities. 
The DMP inference algorithm considerably outperforms the three centrality measures, with a comparable computational cost.

In Fig.~\ref{fig:2} we present the average normalized rank of the true epidemic origin for random regular graphs for the whole range of the transmission probability $\lambda$, for different values of the recovery probability $\mu$, and snapshot times $t_0$. As an estimation for the spreading time $t_0$, we take the one maximizing the ``partition functiom'' $Z(t)=\sum_{i}e^{-E(i,t)}$. The distribution of the estimated time is concentrated at the true spreading time $t_0$. We find that for different values of $\mu$, DMP inference always outperforms the centrality measures (see, e.g., case (a)), except in a special case (b) ($\mu=1$, corresponding to the deterministic recovery), in a range of $0.3<\lambda<0.58$ where Jordan center is a better estimation. In other cases, however, Jordan centrality is less performant. Note that for $\mu<1$ Jordan centrality does not distinguish between recovered and infected nodes, which partly explains its rather bad performance in that case. The Fig.~\ref{fig:2}~(c) shows the dependence on the spreading time $t_0$ for fixed values of $\lambda$ and $\mu$. Note that DMP remains efficient even for relatively large $t_0$, when the centrality algorithms fail to make a prediction.

Importantly, in some range of parameters, the average normalized rank of the true epidemic origin is not so close to zero (note that the value $1/2$ of the normalized rank corresponds to a random guess of patient zero among all the infected or recovered nodes). The problem of estimating the epidemic origin with a good precision is very hard in these regions. In some cases the information about the epidemic origin was lost during the spreading process. For instance for $\lambda>\lambda_c=\mu/(c-2+\mu)$~\cite{Newman02} the epidemic percolates at large times $t_{0}\gg\log_c{N}$;
 then the information about the epidemic origin is lost. On the other hand for $t_{0}<\log_c{N}$, the epidemic is confined to a tree network and in this case the inference of the origin is easier, cf. Fig.~\ref{fig:2}~(c). In Fig.~\ref{fig:2}, (a) and (b), we mostly focused on the intermediate case $t_{0}\approx\log_c{N}$.

\begin{figure*}[!ht]
\vspace{5mm}
\begin{center}
\includegraphics[width=1.96\columnwidth]{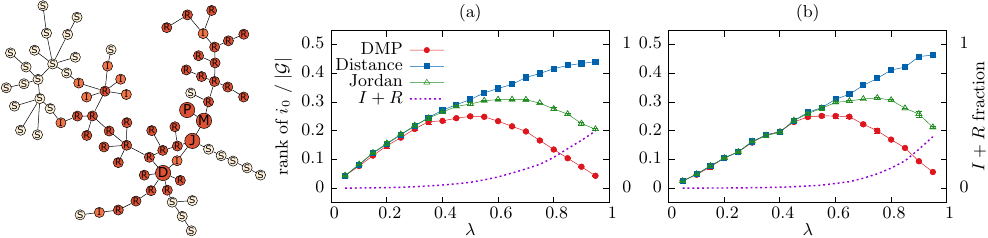}
\caption{(color online) Left: An instance of inference problem on the scale-free graph with average degree $\langle c \rangle=5/3$ and $N=77$. The epidemic is generated for $\lambda=0.7$ and $\mu=0.5$. The snapshot is represented at time $t_{0}=10$. The true patient zero is labeled by $P$, the best-ranked nodes for DMP ($M$), Jordan ($J$) and distance ($D$) centralities are at distances 1, 2 and 4 from $P$, correspondingly. Right: Average rank of the true epidemic origin on scale-free graphs of size  $N\simeq1000$, generated according to the Pareto distribution with shape parameter $\alpha=0.25$, and minimum value parameter $k=1$, average degree $\langle c \rangle=5/3$. Each data point is averaged over $3000$ instances. The snapshot time $t_{0}$ (assumed to be known) and recovery probability $\mu$ are: (a) $t_{0}=10$, $\mu=0.5$ and (b) $t_{0}=10$, $\mu=1$. The dotted line shows the average fraction of nodes that were infected or recovered in the snapshot, $|{\cal G}|/N$, we use this number to normalize the ranks of the epidemic origin.}
\label{fig:PL}
\end{center}
\vspace{-5mm}
\end{figure*}

We also present the results for other families of random networks, that can be qualitatively more relevant for applications. In Fig. \ref{fig:ER} we plot the inference results for the connected component of Erd\"os-R\'enyi graphs of size $N\simeq1000$ with average degree $\langle c \rangle = 4$. Fig.~\ref{fig:PL} shows the corresponding results for the connected component of scale-free networks, that are prototype to the real-world social networks, of size $N\simeq1000$, generated to have the Pareto degree distribution with a shape parameter $\alpha=0.25$, and minimum value parameter $k=1$, with a probability distribution function $P(x)=\alpha k^{\alpha} x^{-1-\alpha}$, defined for $x>k$. For both networks, the DMP algorithm considerably outperforms Jordan and distance centralities. In our opinion the systematic comparison presented here is a good test-bed for comparing and improving algorithms.

\begin{figure}[ptb]
\begin{center}
\includegraphics[width=0.7\columnwidth]{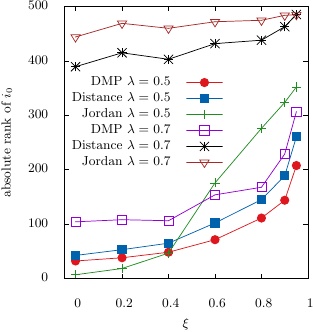}
\vspace{-1mm}
\caption{(color online) Performance of algorithms in the case of an incomplete snapshot. Data is presented for a random contact network of size $N=1000$, degree $c=4$. Recovery probability $\mu=1$, transmission probability $\lambda=0.5$ and $\lambda=0.7$, only the state  of a fraction $1-\xi$ of nodes is observed at time $t_0=10$, assumed to be known. The rank (averaged over $1000$ instances) of the true epidemic origin obtained with our DMP inference algorithm is compared to the distance and Jordan centralities.}
\label{fig:5}
\end{center}
\vspace{-5mm}
\end{figure}

We now show the performance of our algorithm in the case where the snapshot is incomplete: a fraction $\xi$ of nodes is not observed. We compare it to the generalizations of Jordan and distance centralities to this case that we propose in appendix~\ref{app:centralities}. The idea behind this generalization consists in a careful construction of a connected component of infected, recovered and undefined nodes, for which the centrality algorithms can be applied. Fig.~\ref{fig:5} gives the average rank of the true epidemic origin. It shows that, with incomplete snapshots, the DMP inference algorithm outperforms both centralities even in the case where for complete snapshots the Jordan centrality was better. This observed robustness of DMP is a very useful property.

\begin{figure}[ptb]
\begin{center}
\includegraphics[width=\columnwidth]{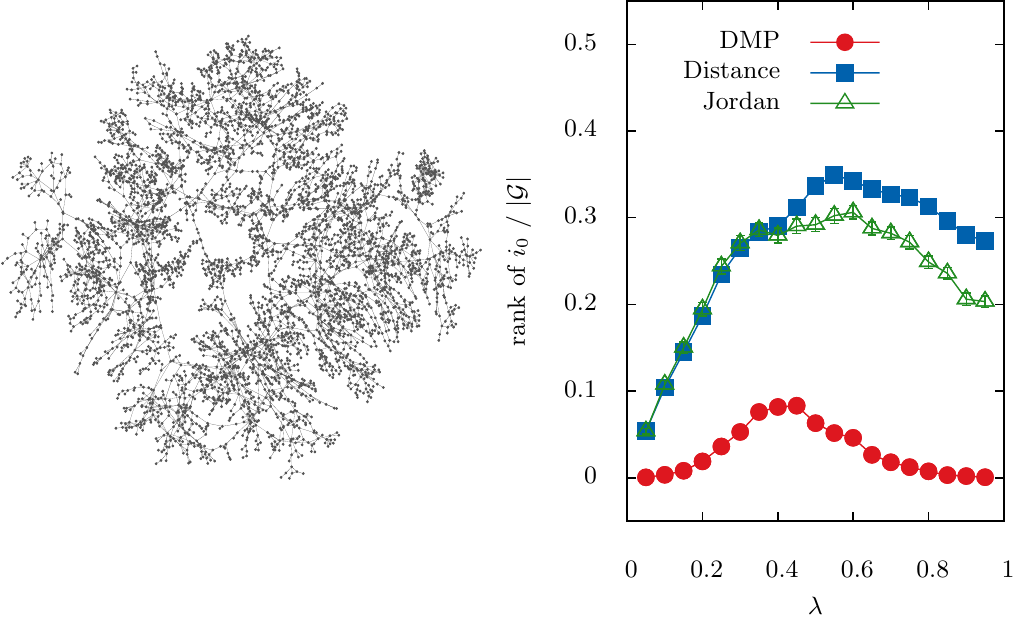}
\vspace{-1mm}
\caption{(color online) Left: A representation of the topology of the U.S. West-Coast power grid network, generated with Gephi. Right: Normalized rank (averaged over 1000 instances) of the true epidemic origin for epidemic spreading with $\mu=0.5$ and all nodes observed at time $t_0=10$, on the power grid network. DMP inference is significantly better than inference based on distance and Jordan centralities.}
\label{fig:5bis}
\end{center}
\vspace{-5mm}
\end{figure}

In order to illustrate the method on non-randomly generated network, we studied the performance of DMP for synthetic data on a real network of the U.S. West-Coast power grid which contains $N=4941$ nodes with a mean degree $\langle c \rangle = 2.67$ and a maximum degree $19$ \cite{WattsStrogatz98}, also considered as an application to the patient zero problem in \cite{ShahZaman10}. Our aim here is not to study any problem relevant to the power grid itself, but we use this well-documented network in order to test how our algorithm performs on a network that is not random. This network is in fact a widely used example of a real network with the small-world property, having a right-skewed degree distribution, and is quite different with respect to an Erd\"os-R\'enyi random graph of the same size and mean degree: its measure of cliquishness, the clustering coefficient $C=0.08$, is much bigger than the transitivity of a corresponding random graph $C_{\text{rand}}=0.005$ \cite{WattsStrogatz98,Newman03}. The results are reported in Fig.~\ref{fig:5bis}: we see that the algorithm works well and DMP estimator gives better prediction for all range of $\lambda$.

Our algorithm is based on an approximate form of Bayesian optimal inference. There are two possible sources of sub-optimality on real networks: first, the fact that the message-passing equations may lead to errors on loopy graphs; and second, the mean-field-like approximation (\ref{eq:CostFunction}) of the joint probability distribution. We have observed that taking into account the two-point correlation in this approximation does not lead to any improvement in our results. It would be interesting to search for better approximations of the likelihood on a general graph.

\begin{figure*}[!ht]
\begin{center}
\includegraphics[width=1.58\columnwidth]{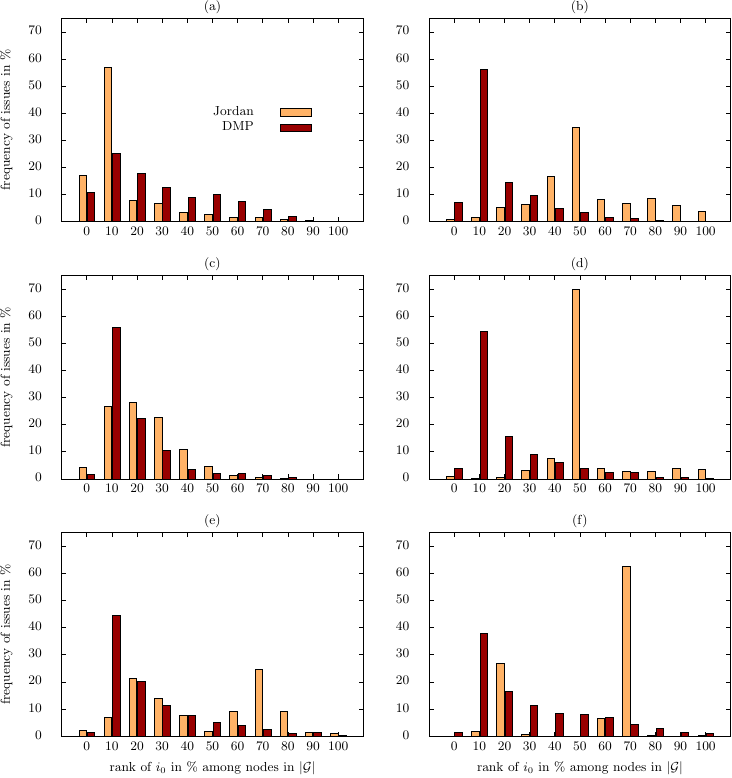}
\caption{(color online) Distribution of inferred rank of the epidemic origin measured over the graph $\mathcal{G}$ for Jordan centrality estimator (light brown) and DMP estimator (dark brown) with known spreading time on regular random graphs of degree $c=4$: (a) $\xi=0,\lambda=0.5$, (b) $\xi=0,\lambda=0.7$, (c) $\xi=0.5,\lambda=0.5$, (d) $\xi=0.5,\lambda=0.7$, (e) $\xi=0.9,\lambda=0.5$, (f) $\xi=0.9,\lambda=0.7$. The average is performed over 500 instances.}
\vspace{-0.5cm}
\label{fig:multihist}
\end{center}
\end{figure*}   

\section{Conclusion}

The approximate solution of dynamics of the SIR model in terms of message-passing equations allowed us to develop an efficient probabilistic algorithm for detecting patient zero. Compared to existing algorithms, it generically (except for a narrow range of parameters) provides an improved estimate for the source of infectious outbreak. It also performs well when the snapshot sees only a part of the network. One superiority of our approach, compared to previous ones, is that it uses efficiently the information about where the epidemic did not spread. As is usual for Bayes inference approaches, our algorithm is versatile and easily amenable to generalizations. Let us mention a few possibilities of extension of our approach, the study of which is left for future work. The present DMP algorithm can be applied to contact networks that evolve in time. The generalization is straightforward, one only needs to encode the dynamics of the network into time-changing transmission probabilities $\lambda_{ij}(t)$ and use the equations (\ref{eq:SIRequations:Sc}-\ref{eq:SIRequations:I}). The SIR model on dynamically changing networks has been already studied using the graph-averaged version of the DMP equations in~\cite{VolzMeyers07,VolzMeyers09}. We anticipate that the DMP equations on a single graph will also be useful for studies where specific experimental data about the changing network, such as those of \cite{StehleVoirin11}, can be used. Our approach can also infer multiple infection sources. In the most straightforward way it would, however, scale exponentially in the number of sources. This can be easily avoided by realizing that with $k$ infection sources, one may do a kind of Monte-Carlo search for their best positions. Another interesting problem for which our approach can be generalized is when the knowledge of the contact network is incomplete.

\begin{acknowledgments}
This work has been supported in part by the EC Grant STAMINA, No. 265496, and by the Grant DySpaN of Triangle de la Physique.
\end{acknowledgments}



\appendix


\section{The centrality algorithms for incomplete snapshots}
\label{app:centralities}

In the case where the state of all the nodes is known at time $t_{0}$, the centrality algorithms work on a connected component $\mathcal{G}$ of infected and recovered nodes. In practice the information is available only for a fraction $1-\xi$ of nodes in the graph $G$. The snapshot $\mathcal{O}(t_{0})$  can then be thought of as a configuration of $(1-\xi)N$ nodes in the states $S$, $I$, $R$ (nodes for which we have the information), and of $\xi N$ randomly located nodes in the unknown state $X$. Now the infected and recovered nodes in general do not form a connected component and are located in several disconnected components, separated by the nodes in the unknown states $X$. Nevertheless, it is clear that not all the $X$-nodes have to be checked as possible candidates to be the actual source of infection. If the cluster of nodes in the $X$ state is surrounded only by the $S$-nodes, this cluster is clearly in the $S$ state itself. Other $X$-nodes in principle are susceptible to be the infection source and thus need to be checked.

We propose the following generalization of centrality algorithms for the $\xi \neq 0$ case. First we construct a connected component composed of all the nodes in the $I$ and $R$ states and clusters of $X$ nodes which are not completely encircled by $S$-nodes. This gives a connected component of $I$, $R$ and $X$ nodes attached together. Since now we have a connected component $\mathcal{G}$, we can run centrality algorithms on it in a usual way. For $\xi=0$ the connected component constructed in this way coincides with a connected component composed of infected and recovered component.

In Fig.~\ref{fig:multihist} we compare the distributions of ranks for DMP and Jordan estimators for different $\xi$ ($\xi=0$, $\xi=0.5$ and $\xi=0.9$) in the special case of ``deterministic'' recovery $\mu=1$ for $\lambda=0.5$ and $\lambda=0.7$. The results are presented for a regular random graph composed of $N=1000$ nodes with connectivity $c=4$, and we take $t_{0}=10$. The plot shows how often the rank of the actual epidemic origin $i_{0}$ is within the value of the corresponding bin ($0 \%$ means exact reconstruction). According to the histogram, in $60 \%$ of cases we manage to locate the true infection source within $10 \%$ of relevant nodes (those situated in $\mathcal{G}$) for $\xi=0$. This number falls to $40 \%$ for $\xi=0.9$, when the states of only $10 \%$ of nodes in the network are known.

We see that although for $\xi=0$ the rank distribution based on the Jordan centrality estimator gives better results (in the case $\lambda=0.5$), it is no longer efficient when the number of unknown nodes gets larger (for all $\xi > 0.4$). The dependence on $\xi$ for the case $\lambda=0.7$ follows the same patterns.


\twocolumngrid
\bibliography{myentries}

\end{document}